\documentclass{llncs} 
\usepackage[utf8]{inputenc}
\usepackage[ngerman,english]{babel}

\usepackage{color}
\usepackage{graphicx,caption,subcaption,amsmath,amssymb,tabularx,array,booktabs} 
\usepackage{setspace}
\usepackage{listings,xcolor}
\usepackage{multirow}
\usepackage{array}
\newcolumntype{P}[1]{>{\raggedright\arraybackslash}p{#1}}

\usepackage{graphicx}
\usepackage{wrapfig}
\usepackage{lscape}
\usepackage{rotating}

\usepackage{listings}
\usepackage{chngcntr}
\AtBeginDocument{\counterwithout{lstlisting}{section}}

\usepackage{floatrow}
\newfloatcommand{capbtabbox}{table}[][\FBwidth]

\newcommand{\nop}[1]{}

\usepackage{booktabs,graphicx}
\usepackage{float}
\usepackage{enumitem}
\usepackage{array}
\newcolumntype{P}[1]{>{\raggedright\arraybackslash}p{#1}}
\usepackage{graphicx}
\usepackage{wrapfig}

\title{Linked Crunchbase: A Linked Data API and RDF Data Set About Innovative Companies}

\author{
Michael Färber
}

\usepackage{todonotes}
\PassOptionsToPackage{hyphens}{url}\usepackage[hidelinks,colorlinks=false, pdfborder={0 0 0}]{hyperref}
\newcommand{\keywords}[1]{\par\addvspace\baselineskip
\noindent\keywordname\enspace\ignorespaces#1}

\institute{Institute AIFB, Karlsruhe Institute of Technology (KIT), Germany\\\email{michael.faerber@kit.edu}
}

\usepackage[none]{hyphenat}
\newcolumntype{P}[1]{>{\centering\arraybackslash}p{#1}}
\newcolumntype{R}[1]{>{\raggedleft\arraybackslash}p{#1}}
\newcolumntype{L}[1]{>{\raggedright\arraybackslash}p{#1}}
\newcolumntype{C}[1]{>{\centering\arraybackslash}p{#1}}

\begin{document}
\maketitle

\begin{abstract}
Crunchbase is an online platform collecting information about startups and technology companies, including attributes and relations of companies, people, and investments. Data contained in Crunchbase is, to a large extent, not available elsewhere, making Crunchbase to a unique data source. 
In this paper, we present how to bring Crunchbase to the Web of Data so that its data can be used in the machine-readable RDF format by anyone on the Web. 
First, we give insights into how we developed and hosted a Linked Data API for Crunchbase and how \texttt{sameAs} links to other data sources are integrated. Then, we present our method for crawling RDF data based on this API to build a custom Crunchbase RDF knowledge graph. We created an RDF data set with over 347 million triples, including 781k people, 659k organizations, and 343k investments. 
Our Crunchbase Linked Data API is available online  at \url{http://linked-crunchbase.org}. 
\end{abstract}

\section{Introduction}

Crunchbase\footnote{See \url{https://www.crunchbase.com/} (Accessed: 18 July 2019).} is an online platform providing information about startups, technology companies, and related entities, such as the key people, the investments they made and received, and the acquisitions they conducted. 
Crunchbase is mainly used by entrepreneurs, investors, and business analysts to look up information for gaining market insights \cite{Skala2019}.\footnote{See \url{https://about.crunchbase.com/partners/advertising-partners/} (Accessed: 18 July 2019).} 
Data samples of Crunchbase have also been used for various research purposes. Dalle et al. \cite{Dalle2017} present about 100 studies that are based on Crunchbase data. Since Crunchbase contains data which is not easily obtainable from other sources (e.g., detailed information about investments and investors), it has been used, for instance, to determine success factors of startup accelerators \cite{Merilainen2016,Dalle2017} and to analyze the gender distribution \cite{Ewens2018} among venture firms and startup founders.\footnote{See also \url{https://news.crunchbase.com/news/announcing-2017-update-crunchbase-women-venture-report/} and \url{https://motherboard.vice.com/en_us/article/nz77zd/female-investors-female-startups-crunchbase-data} (Accessed: 18 July 2019).}

The Crunchbase data is edited by a community: registered users can add and delete entities, as well as add and modify facts about these entities through a browser user interface.
About 3,000 investment companies update the Crunchbase data based on their portfolios.\footnote{See \url{https://support.crunchbase.com/hc/en-us/articles/360009616013-Where-does-Crunchbase-get-their-data-} (Accessed: 18 July 2019).} Consequently, Crunchbase can be regarded as a rich and continuously updated (and therefore up-to-date) knowledge graph. 
Since the Crunchbase data is internally stored as a graph with predefined entity types, attributes, and relations, it is amendable to be modeled in RDF. So far, however, only a JSON REST API has been officially made available to users and no public data dumps have been provided. 

Having the Crunchbase data available via a REST-ful API with RDF as data format and in the form of an RDF data set would enable us to do the following: 
\begin{enumerate}
    \item We can execute complex SPARQL queries against the Crunchbase RDF data set which go beyond plain keyword-based information retrieval and beyond queries about single entities \cite{Faerber2018Crunchbase}.
    \item We can extend other RDF data sets with Crunchbase data more easily as RDF is used as common data model. This allows the development of novel applications, such as intelligent job search engines \cite{Mochol:2007}.
    \item We can apply knowledge discovery and data mining methods to the Crunchbase data which has not been possible due to the prior incompatible data formats and missing interlinkage with other data sets. For example, by establishing \texttt{sameAs} links between Crunchbase entities and DBpedia entities, widely used text annotation methods developed for DBpedia and Wikipedia can be applied out-of-the-box \cite{Faerber2016}.
\end{enumerate}

All linked data wrappers which have been developed so far for Crunchbase do not work any more due to updates on the underlying Crunchbase API and other reasons \cite{Faerber2018Crunchbase}. 
The Crunchbase data sets used and published so far (e.g., for economic studies) 
have not been represented in RDF or 
only cover small parts of the whole Crunchbase database.\footnote{For instance, the Crunchbase data set of \cite{Faerber2018Crunchbase} does not contain detailed information about investments, investors, funding rounds, news articles, etc.} Note also that the licensing model of Crunchbase has been changed, leading to some limitations.\footnote{Until 2016, the Crunchbase data had been licensed partly under Creative Commons Attribution-NonCommercial License 4.0 (CC-BY-NC) and partly under Creative Commons Attribution License 4.0 (CC-BY), independent how it had been provided. Nowadays, Crunchbase only allows sharing its data based on a commercial model (see \url{https://about.crunchbase.com/docs/terms-of-service/} [Accessed: 18 July 2019]).} Newly created Crunchbase data sets may not be shared free of charge any longer. Knowing how to crawl a Linked Data API like our Crunchbase Linked Data API for research purposes (see examples of in-depth data analysis in Sec.~\ref{sec:usage}) is, in our view, still worth investigation. Furthermore, the commercial licensing model for non-research purposes makes it especially attractive to know whether entities in commonly-used linked data sources, such as DBpedia, also occur in Crunchbase (known via \texttt{sameAs} relations). In these cases only Crunchbase data about those entities needs to be acquired.

\begin{table*}[tb]
    \centering
        \caption{Links to resources.}
    \label{tab:important-links}
    \begin{small} 
    \begin{tabular}{p{3cm} p{9cm}}
    \toprule
    Description & URI \\
    \midrule
Linked Data API:
& \url{http://linked-crunchbase.org/} \\
Ontology: & \url{http://linked-crunchbase.org/ontology.owl} \\
Source code:
& \url{https://github.com/michaelfaerber/linked-crunchbase/} \\
    \bottomrule
    \end{tabular}
    \end{small}
\end{table*}

\begin{figure*}[tb]
    \centering
    \includegraphics[width=\linewidth]{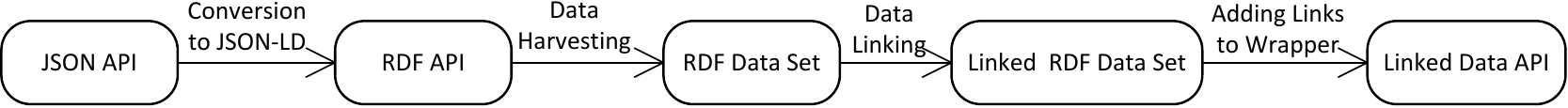}
    \caption{Schematic view of the steps taken to create a Linked Data version of the Crunchbase API.}
    \label{fig:flowchart}
\end{figure*}

Overall, we make the following contributions in this paper:
\begin{itemize}
\item Based on Crunchbase as a use case, we provide a process-oriented description of creating a Linked Data API which provides both JSON-LD and N-Triples serializations (see Fig.~\ref{fig:flowchart}). The Linked Data API is a wrapper around the existing API of Crunchbase. 
  Our implementation of the Linked Data API and a deployed version of it are available online (see Table~\ref{tab:important-links} and \url{https://doi.org/10.5281/zenodo.2160359}). 
Previous versions of our Crunchbase Linked Data API have already been applied in several use cases (see Sec.~\ref{sec:usage}). 
 \item We present a methodology for creating an up-to-date RDF data set of Crunchbase using our Crunchbase Linked Data API.
 This RDF data set can be used for a variety of use cases, such as monitoring companies in the news or predicting the success of startups. Due to licensing issues, our data set obtained in the experiments cannot be shared; however, by making use of our detailed guidelines, it can be easily reproduced.\footnote{See \url{http://linked-crunchbase.org} (Accessed: 18 July 2019).}  
Previous versions of our Crunchbase RDF data set have been used by others for data integration efforts (see Sec.~\ref{sec:usage}). Non-semantically-structured Crunchbase data sets have been used for exploratory data analysis.
\end{itemize}

The rest of our paper is structured as follows: 
In Section~\ref{sec:crunchbase-api}, we present our Linked Data API for Crunchbase, which is designed as a wrapper around the official Crunchbase REST API. 
In Section~\ref{sec:crunchbase-dump}, we give insights into our Crunchbase RDF data set. 
After describing the 
usage of the Linked Data API and the crawled RDF data in Section~\ref{sec:usage}, we conclude with Section~\ref{sec:conclusions}.

\section{The Crunchbase Linked Data API}
\label{sec:crunchbase-api}

\begin{figure*}[tb]
 \centering
  \includegraphics[width=.75\linewidth]{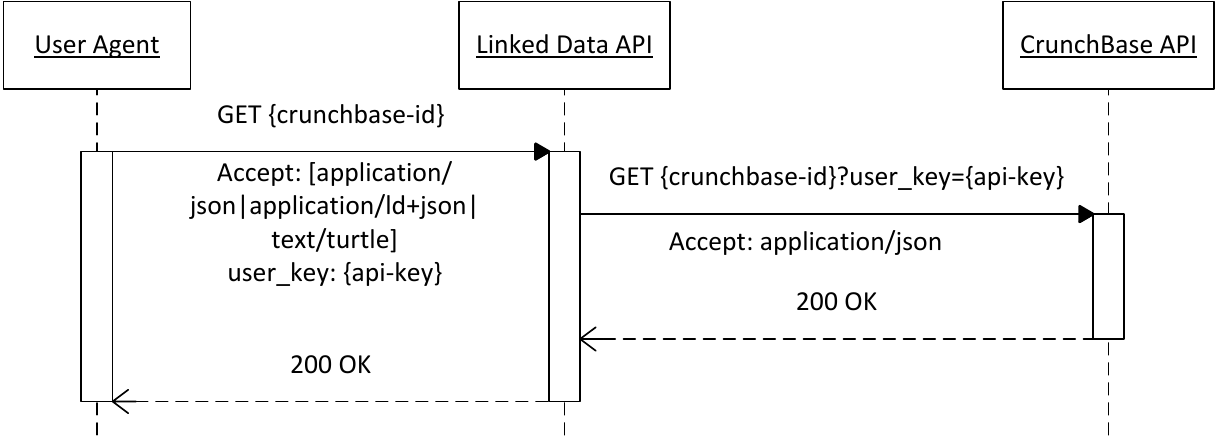} 
 \caption{UML sequence diagram illustrating the use of the wrapper.  The wrapper supports different representations via content negotiation. The API key is passed to the wrapper via an {\tt Authorization} header, and passed from the wrapper to the Crunchbase API as URI parameter.}
 \label{fig:wrapper}
\end{figure*}

We now describe the development of our Crunchbase Linked Data API. 
Fig.~\ref{fig:wrapper} shows the basic workflow for using this API. We distinguish between the following steps:
\begin{enumerate}
 \item A user application (e.g., the data integration system Linked Data-Fu \cite{Stadtmuller:2013}) calls the Crunchbase Linked Data API via a HTTP GET 
 request. The request contains the URI, the requested content type, and the Crunchbase API user key.\footnote{An example API call with cURL is
 \texttt{curl -v -H "Accept:text/turtle" --header "Authorization: Basic \{Base64-encoded key\}" \url{http://linked-crunchbase.org/api/organizations/facebook}}.}
 \item The Linked Data API (implemented as a servlet) takes the HTTP request and calls the official Crunchbase REST API using the specified information.
 \item The Linked Data API receives the data from the Crunchbase REST API and transforms it into one of the provided content types. \texttt{sameAs} links to DBpedia entities are included as far as possible. 
 \item The user application receives the data from the Linked Data API and further processes the data.
\end{enumerate}

Our Crunchbase Linked Data API is capable of providing the following three content types:
\begin{enumerate}
 \item \texttt{JSON (application/json)}: The official Crunchbase REST API provides data in JSON format. For JSON responses, we forward the data retrieved from the Crunchbase REST API without any modifications.
 \item \texttt{JSON-LD (application/ld+json)}:
 For providing data via our Crunchbase Linked Data API as JSON-LD, we restructure the JSON file retrieved from the official Crunchbase API. The main restructuring steps involve removal of metadata and addition of namespaces. Additionally, Crunchbase encapsulates properties (e.g., date of birth of a person), relationships (e.g., acquisitions of a company) and items in lists. 
To avoid blank nodes, we removed the list structure. 
 \item \texttt{RDF/N-Triples (text/turtle)}: We also provide data in the form of N-Triples, a subset of the Turtle syntax for RDF. This is one of the widely used formats in current Semantic Web systems. 
\end{enumerate}

Since we provide the Crunchbase Linked Data API as a third-party tool on top of the Crunchbase REST API (currently in version 3.1), the Crunchbase Linked Data API needs to be modified as soon as the Crunchbase API changes. This is ensured by a process of monitoring the Crunchbase mailing list and by an automated monitoring of the Crunchbase' API documentation website. In the past, this process allowed us to update our Linked Data API when the Crunchbase API changed from version 3.0 to 3.1.

\subsection{API Authorization}

Since the official Crunchbase API is only accessible with an API key, users of the Crunchbase Linked Data API also need to provide a valid API key for requesting data.
When using the Crunchbase JSON API, the key is passed via a parameter in the URI. However, applying this method to the Crunchbase Linked Data API, the API key would be part of the identifier and therefore public for everyone. To resolve this issue,
user agents can pass the API key through the \texttt{Authorization} header field.\footnote{We use the \texttt{Basic Authentication} method. The key is stored in the ``user'' field; the ``password'' field remains empty.} Our approach allows a neat integration of the Crunchbase Linked Data API with other services and frameworks, since we use standard web technologies and the URIs do not need to be modified due to authorization. 

If no API key is given, our Crunchbase Linked Data API still returns RDF data. This approach ensures that all URIs provided by the Crunchbase Linked Data API are dereferencable and can be requested by anyone on the Web. Our Linked Data API is therefore also visible to and partly usable by users who follow a link to our API, but do not possess an API key. 
Note, however, that the Crunchbase data cannot be freely shared due to the changed licensing model. Thus, 
our Crunchbase Linked Data API merely returns license-free \texttt{owl:sameAs} links concerning the requested resource in case no API key is provided. 

\begin{table}[tb]
 \caption{URI design for the Crunchbase Linked Data API.} 
 \label{tab:uri-design}
 \begin{footnotesize}
 \begin{tabular}{p{6.9cm}p{5cm}}
 \toprule
 URI Template & Description \\
 \midrule
 \texttt{/}& Index page \\
 \texttt{/api/} & Base for every request \\
 \texttt{/api/\{entity-type\}} & Returns all instances of the \texttt{entity-type} (e.g., \texttt{organizations}) in Crunchbase \\
 \texttt{/api/\{entity-type\}/\{permalink\}} & Returns information about an entity denoted as \texttt{permalink} (e.g., \texttt{facebook}) \\
  \texttt{/api/\{entity-type\}/\{permalink\}/\{relation\}} & Returns information about a \texttt{relation} (e.g., \texttt{acquisitions}) of an entity (e.g., \texttt{facebook}) \\
 \bottomrule
 \end{tabular}
 \end{footnotesize}
\end{table}

\subsection{URI Schema} 

Table \ref{tab:uri-design} shows the URI design for accessing the Linked Data API.
Since the URIs for the official Crunchbase API
and the Linked Data API are designed in the same way, every request sent to the official Crunchbase API can also be sent to our Crunchbase Linked Data API.

\subsection{Used Schema} 
\label{sec:schema}

\begin{figure*}[tb]
 \centering
 \includegraphics[width=0.99\linewidth]{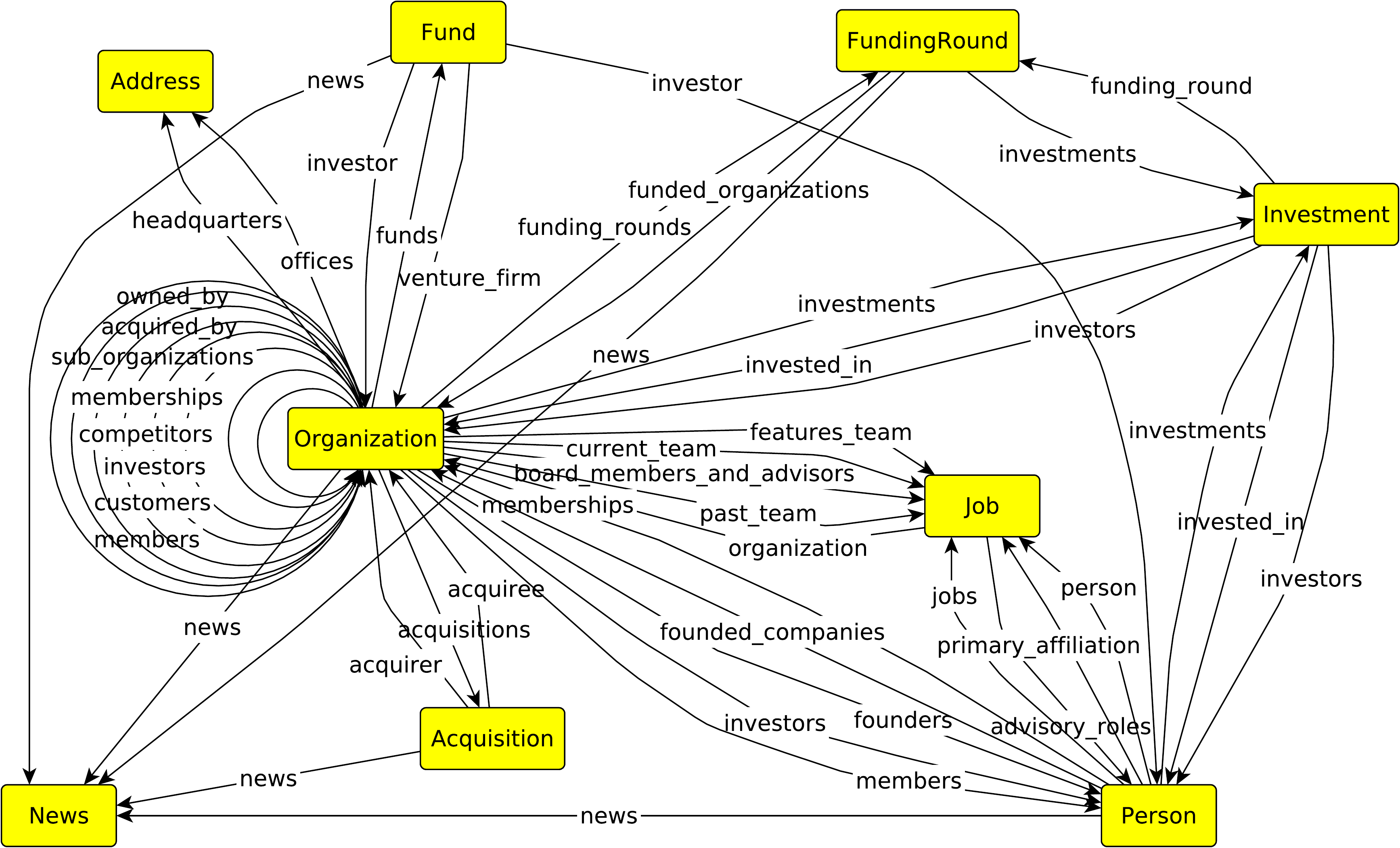} 
 \caption{Subset of the classes and object properties of the Crunchbase schema.}
 \label{fig:main-classes}
\end{figure*}

For the Crunchbase Linked Data API, the data model of the official Crunchbase REST API is reused and only slightly modified. All entity types and the set of possible attributes and relations between entities remain. 
Fig.~\ref{fig:main-classes} illustrates some classes and relations of the used schema.\footnote{A complete schema overview is given online and not shown here due to the large size.}

The schema of the Linked Data API is described in a dereferenceable OWL file, which is provided on our Linked Data API entry page. We enriched our ontology with Vocabulary-of-a-Friend (VOAF)\footnote{See \url{http://lov.okfn.org/vocommons/voaf} (Accessed: 18 July 2019).} descriptors
in order to link our ontology to other vocabularies and to introduce the 
vocabulary to the Linking Open Data community.\footnote{See \url{https://www.w3.org/wiki/SweoIG/TaskForces/CommunityProjects/LinkingOpenData} (Accessed: 18 July 2019).} 

We can outline further characteristics of the data model used in our Crunchbase Linked Data API:
\begin{enumerate}
 \item Relations are often stored in a duplicated fashion to provide easier querying capabilities for users and machines. For instance, 'which organization acquired which company (with which amount)' is represented via the auxiliary entity type \texttt{:Acquisition}. However, users and machines can also obtain the acquisition information by using the \texttt{:acquisitions} relation of \texttt{:Organization}. A similar phenomenon can be observed for investments and other n-ary relations. 
\item Noteworthy is also the possibility to model uncertainty for date values in Crunchbase.
The uncertainty value is stored as a decimal ranging from 0 (complete unknown/unsure) to 7 (very confident/knowing the exact date). Based on this encoding, property values stored as strings can easily be converted to the XML schema definition (XSD) format\footnote{See \url{https://www.w3.org/2001/XMLSchema} (Accessed: 18 July 2019).} such as \texttt{xsd:date} if they are valid.
\end{enumerate}

Linked Data is based on the best practices of using existing vocabularies and linking instances, classes, and properties between data sources in the Linked Open Data (LOD) cloud. As Crunchbase covers mainly domain-dependent information, such as information about investments, funding rounds, and acquisitions, we could not find a suitable external vocabulary for Crunchbase. Therefore we decided to use the vocabulary which we already used for creating a preliminary Crunchbase RDF data set in 2015~\cite{Faerber2018Crunchbase} and to update the links of all entity types and properties referring to \texttt{schema.org}.
We created 32
\texttt{equivalentProperty}, 
16
\texttt{subProper\allowbreak tyOf}, 
8~\texttt{subClassOf}, and 
7~\texttt{equivalentClass} links. 
Table \ref{table:schema-org-mappings} shows some examples of entity types which are linked to \texttt{schema.org}. The list of all mappings is provided in our created OWL file.

\begin{table}[tb]
 \caption{Examples of mappings between Crunchbase 
 and \texttt{schema.org} entity \\\hspace{\textwidth} types. We use \texttt{cbw} as prefix for \url{http://ontologycentral.com/2010/05/cb/vocab\#}.}
 \label{table:schema-org-mappings}
  \centering
 \begin{small}
 \begin{tabular}{ll}
\hline
Crunchbase entity type ~& \texttt{schema.org} entity type \\
\hline
\texttt{cbw:Address} & \texttt{schema:Place} \\
\texttt{cbw:Image} & \texttt{schema:ImageObject} \\
\texttt{cbw:News} & \texttt{schema:NewsArticle} \\
\texttt{cbw:Organization} & \texttt{schema:Organization} \\
\texttt{cbw:Person} & \texttt{schema:Person} \\
\texttt{cbw:Video} & \texttt{schema:VideoObject} \\
\texttt{cbw:Website} & \texttt{schema:WebSite} \\
\hline
 \end{tabular}
 \end{small}
\end{table}

\subsection{Linking Crunchbase to DBpedia} 
\label{sec:linking-entities}

We created mappings in the form of \texttt{owl:sameAs} statements between Crunchbase entities and the corresponding DBpedia entities.\footnote{We also experimented with mappings to Wikidata. However, we obtained less mappings than with DBpedia.} We provide these mappings as an RDF document for download and for specific entities as additional triples in the Crunchbase Linked Data API result.

\subsubsection{Organization Mappings}
For each organization in Crunchbase, we checked whether we could find a DBpedia entity which possesses the same homepage domain. 
Concretely, we compared the value of the entity's \texttt{homepage} property in Crunchbase with the value of the entity's  \texttt{foaf:homepage} property in DBpedia. 
For a robust string comparison, we only considered the fully qualified domain name (FQDN) instead of the full URLs. If there was a match, the entity pair was added to the mapping list. 
In total, we obtained 
1,155
mappings for all 
659k
Crunchbase organization entities.
Thus, the recall of the mappings is very low. 
This might be due to several reasons. 
The overlap between Crunchbase entities and DBpedia entities is generally quite low. 
Crunchbase deals with entities and facts in the business world. 
Wikipedia, -- as the data source for DBpedia -- allows anyone to edit pages and create new ones concerning any general knowledge, but applies quite strict rules on what to include into and keep in Wikipedia.\footnote{See the Wikipedia's notability guidelines at \url{https://en.wikipedia.org/wiki/Wikipedia:Notability} (Accessed: 18 July 2019).} Bots and Wikipedia contributors delete new entities if they do not deem them to be of general public interest. As a result, Wikipedia does not contain many startups. 
Another reason for the low recall is that \texttt{foaf:homepage} properties in DBpedia often are either missing for represented organizations or have noisy values. 

To evaluate the precision of the gained \texttt{owl:sameAs} links, we manually evaluated 100 randomly chosen \texttt{owl:sameAs} triples of Crunchbase organization entities by checking the corresponding websites. 89 of 100 were judged to be correct. Five triples were incorrect. In six cases, no final judgment could be made due to insufficient information.

\subsubsection{People Mappings}

People entities require a higher effort for mapping than organization entities. Using just the given name and family name leads to a very high rate 
of false positives, since a lot of people have the same names, but correspond to different entities (e.g., Brian Ray, who is the CEO of Link Labs on Crunchbase, but a musician on DBpedia).\footnote{See \url{https://www.crunchbase.com/person/brian-ray\#/entity} and \url{http://dbpedia.org/page/Brian\_Ray} (Accessed: 18 July 2019).} 
To avoid this type of error, we took both the name and the birthday of a person into account for the mapping. This resulted in 1,294 mappings. Although this accounts for a small fraction of all people represented in Crunchbase, these 1,294 people are quite famous or popular, making these mappings valuable.

To evaluate the accuracy of this mapping strategy, 
we randomly picked 100 Crunchbase person entities 
and for each entity verified via manual investigation on Wikipedia whether there is a corresponding entity in DBpedia. 
We came to the following conclusions: 
95 out of the 100 (95\%) Crunchbase people entities are correctly linked to DBpedia.
For the remaining five people, not enough information was provided. Thus, no definite assessment about the linking could be done. However, no contradictory information was visible either.

\section{The Crunchbase RDF Data Set}
\label{sec:crunchbase-dump}

\subsection{Creating the Data Set} 
\label{sec:creating-data-set}

We built an RDF data set containing all data from Crunchbase  
by taking the following steps: 
\begin{enumerate}
\item Crunchbase provides daily updated CSV files on its website\footnote{See \url{https://data.crunchbase.com/docs/daily-csv-export} (Accessed: 18 July 2019).} with 
some information about entities of specific entity types in Crunchbase. 
We generated URIs out of this data and crawled all Crunchbase data available in this way 
using our Crunchbase Linked Data API. 
We followed the \texttt{next\_page\_URI} in the retrieved RDF documents since the Crunchbase API spreads information across multiple pages.
The result is a set of entity summaries. 
 \item To obtain all attributes and relations in Crunchbase, 
we made requests against our Crunchbase Linked Data API using the URIs in the \texttt{api\_path} field of the summary data obtained in the previous step. 
 \item Since the Crunchbase API lists only eight entities in object position 
 when requesting information about a specific relation, we crawled every relation of any entity separately in case the relation has more than eight objects. 
\end{enumerate}
Note that the API calls take time, because the official Crunchbase REST API, which is used by the Crunchbase Linked Data API, has a limit of 1 request per second. 
Due to the current Crunchbase licensing model, we cannot make the resulting RDF data set -- containing 347 million RDF triples -- publicly available. However, step-by-step instructions for developers are provided on our GitHub repository.
Given that a developer has a Crunchbase API key (available for free for research purposes), she can recreate our RDF data set with ease.

{\centering
\begin{minipage}{0.53\textwidth}
   \centering
   \includegraphics[width=\linewidth]{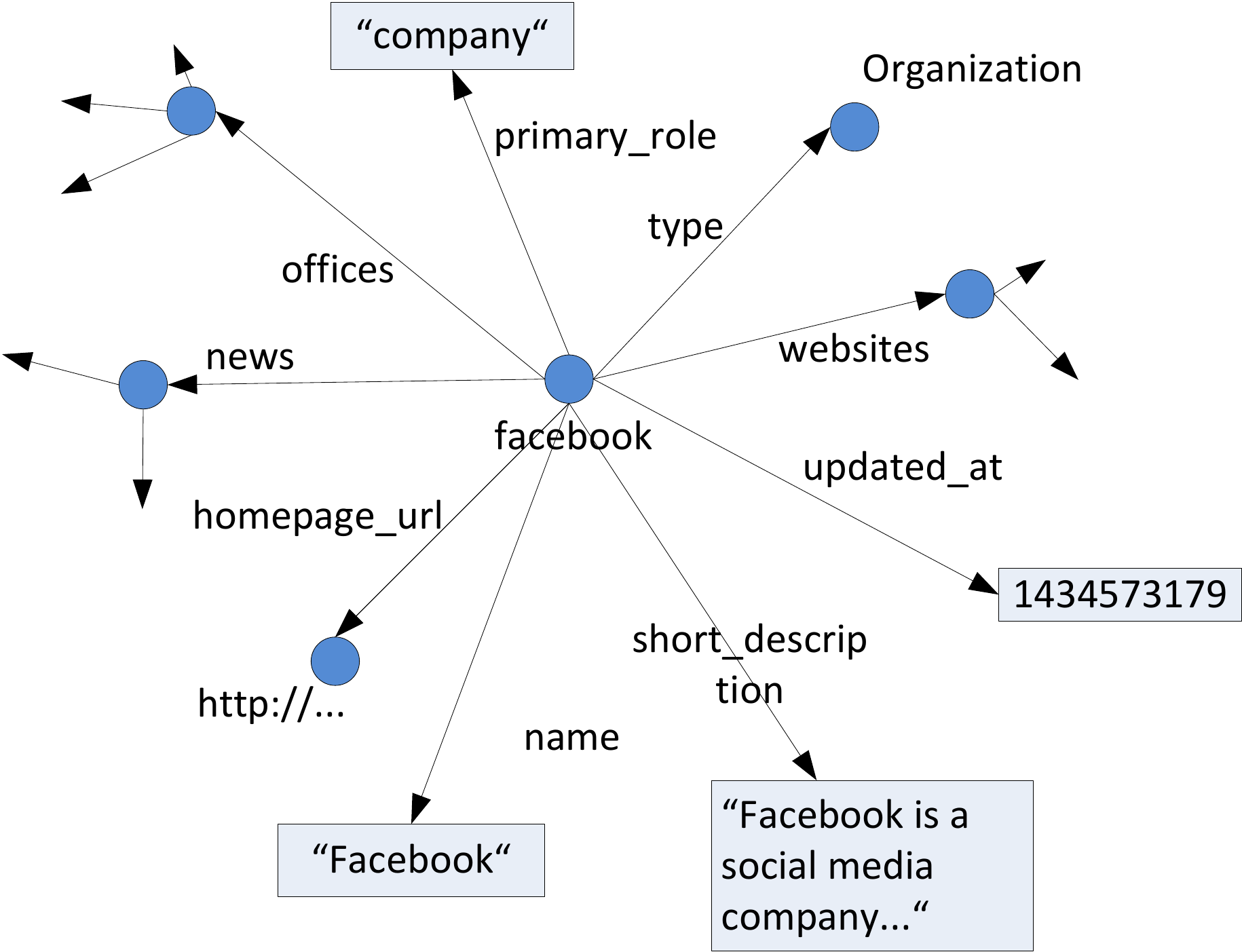}
   \captionof{figure}{Subgraph of our Crunchbase knowledge graph showing information about a company.}
    \label{fig:cb-example}
\end{minipage}
~
\begin{minipage}{0.46\textwidth}
   \centering
   \begin{footnotesize}
\begin{tabular}{lr}
 \toprule
  Entity type & \# Instances \\
  \toprule
  News	& 	5,845,188 \\
  Jobs	&  3,611,335 \\ %
  Websites	& 2,282,952	\\ %
  People 	& 	780,727 \\
  Organizations	& 658,963 \\
  Addresses	& 	447,705 \\
  Investments		& 342,547 \\
  Degrees	& 	276,653 \\
Funding Rounds	& 	222,244 \\
Acquisitions	& 	77,105 \\
IPOs	& 	16,037 \\
Locations	& 	12,211 \\
Funds	& 	9,349 \\
Categories	& 	739 \\
  \bottomrule
 \end{tabular}
 \end{footnotesize}
   \captionof{table}{Distribution of entities among the different entity types in our crawled Crunchbase data set (crawling started in June 2018).}
    \label{tab:distribution-dump}
\end{minipage}
}

\subsection{Data Set Statistics}

Overall, our crawled Crunchbase data set contains entities of 16 entity types, 
39 different object properties, and 101 different data type properties. 
We retrieved 
346,695,771 unique triples in total. Table \ref{tab:distribution-dump} shows the distribution of the entities among the different entity types. Not surprisingly, Crunchbase's main focus is on organizations (including companies) and related entities, such as people, acquisitions, and jobs. News and websites are also well covered due to the affiliation of Crunchbase to TechCrunch.

As outlined in Sec.~\ref{sec:linking-entities}, we created a total of 2,449 \texttt{owl:sameAs} links between Crunchbase entities 
and DBpedia entities. We provide these mappings 
as a separate RDF data set online for further use. 

\begin{lstlisting}[captionpos=b,label={lst:query1},caption={Querying the birth dates of investors stored in Crunchbase via SPARQL.},basicstyle=\scriptsize\ttfamily,frame=single,breaklines=true,float=t]
PREFIX rdf: <http://www.w3.org/1999/02/22-rdf-syntax-ns#>
PREFIX cb: <http://ontologycentral.com/2010/05/cb/vocab#>
PREFIX xsd: <http://www.w3.org/2001/XMLSchema#>

SELECT ?person ?bornon
FROM <http://linked-crunchbase.org>
WHERE {
 ?person rdf:type cb:Person .
 ?person cb:role_investor "true"^^xsd:boolean .
 ?person cb:born_on ?bornon.
}
\end{lstlisting}

\begin{figure}[tb]
\centering 
    \caption{Statistics of our obtained Crunchbase RDF data set.}
    \label{fig:example-visualization}
\begin{subfigure}{0.49\textwidth}
    \includegraphics[width=\linewidth]{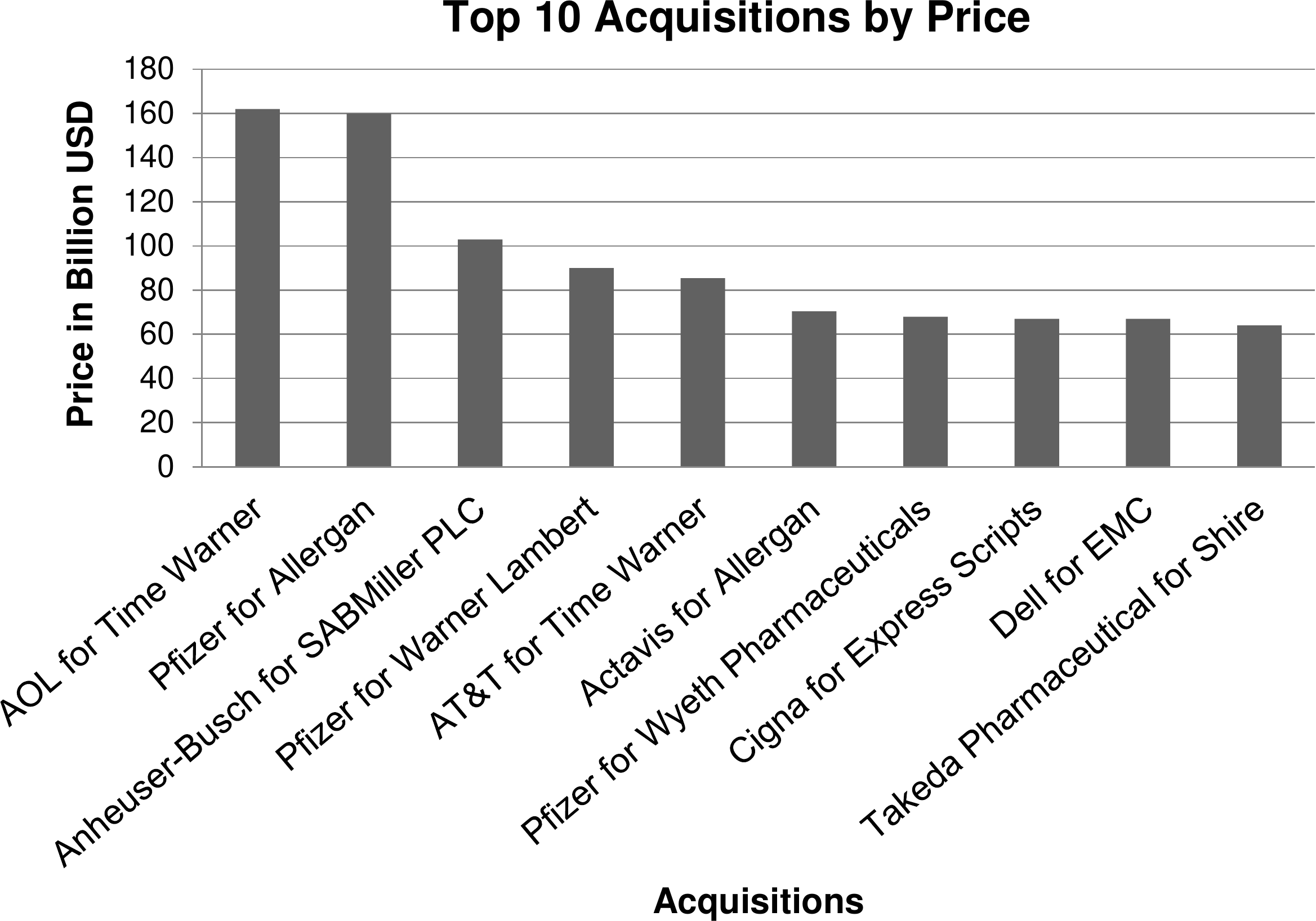}
\end{subfigure}\hfil 
\begin{subfigure}{0.49\textwidth}
\raisebox{0.59cm}{
    \includegraphics[width=\linewidth]{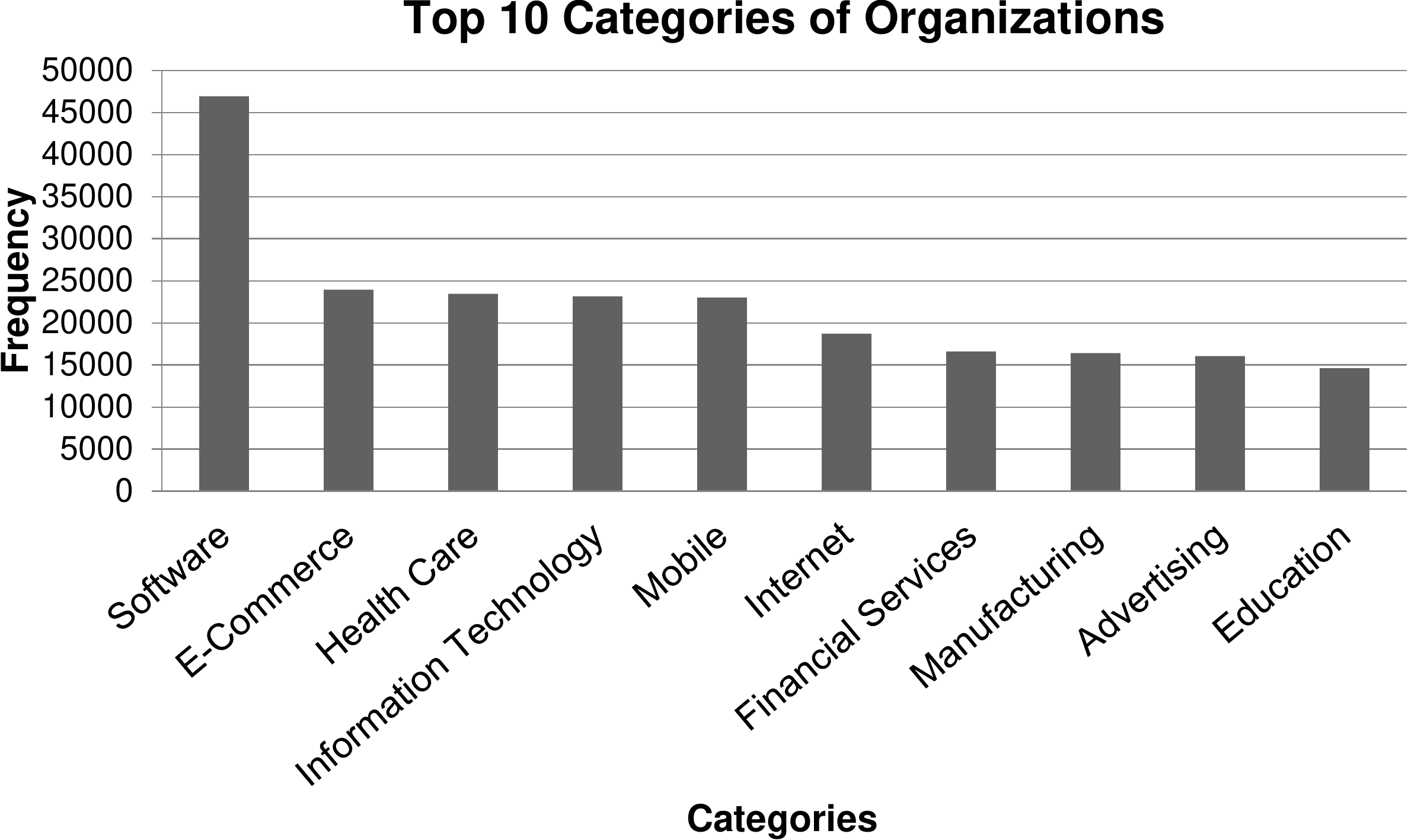}
    }
\end{subfigure}
\\
\vspace*{0.5em}
\begin{subfigure}{0.49\textwidth}
    \includegraphics[width=\linewidth]{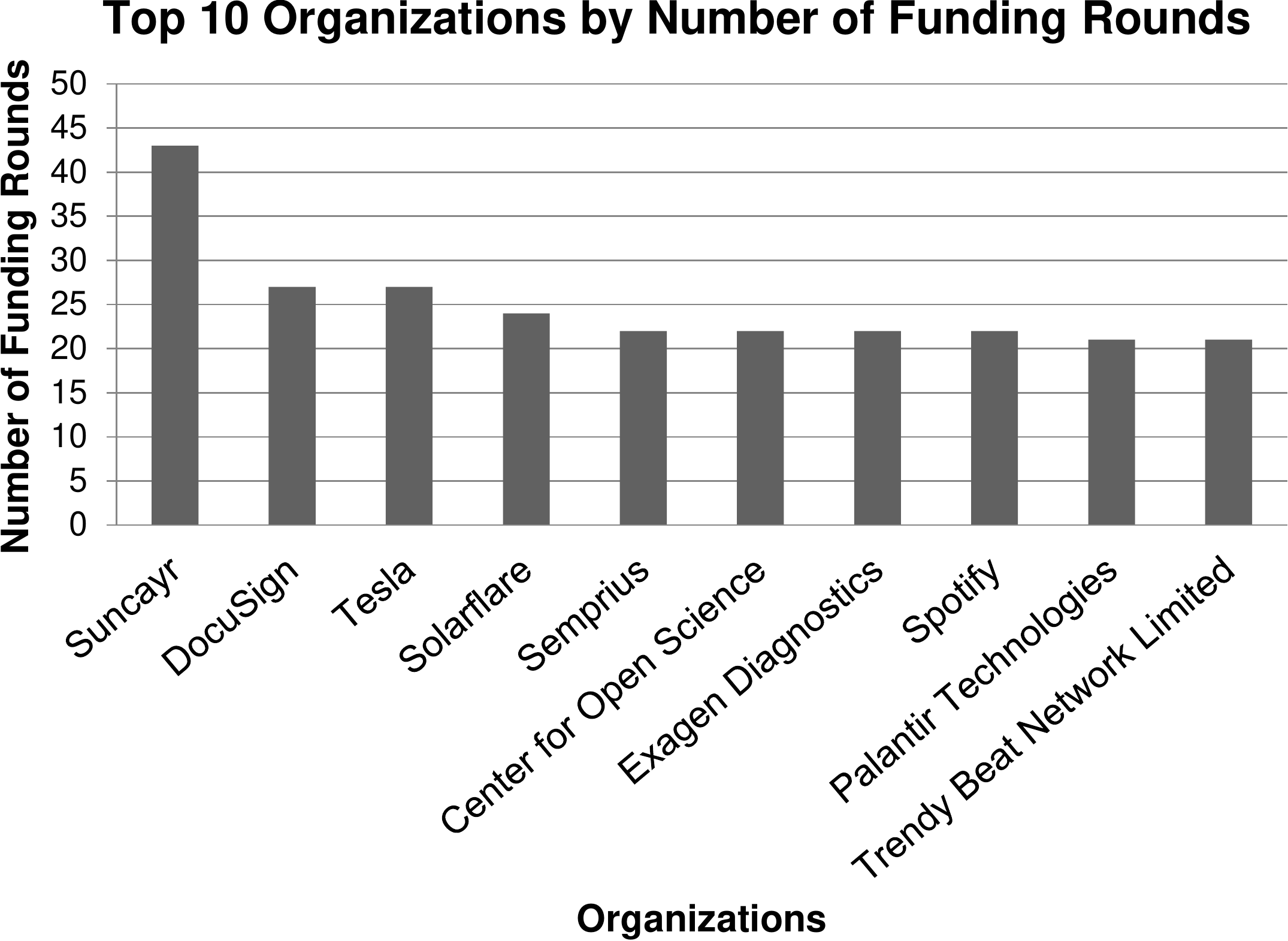}
\end{subfigure}\hfil 
\begin{subfigure}{0.49\textwidth}
\raisebox{0.4cm}{
\includegraphics[width=\linewidth]{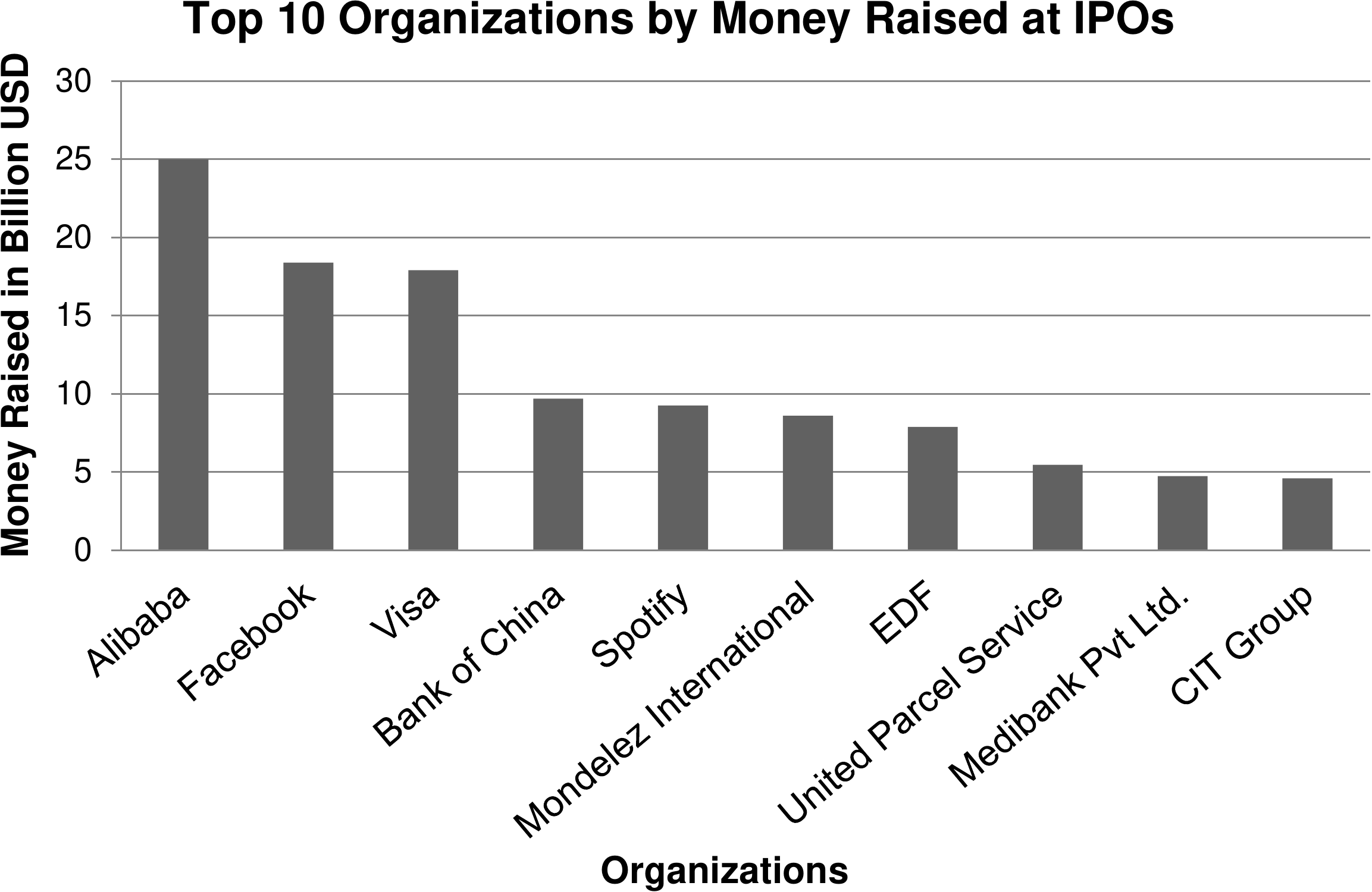}
}
\end{subfigure}
\\
\vspace*{0.5em}
\begin{subfigure}{0.49\textwidth}
    \includegraphics[width=\linewidth]{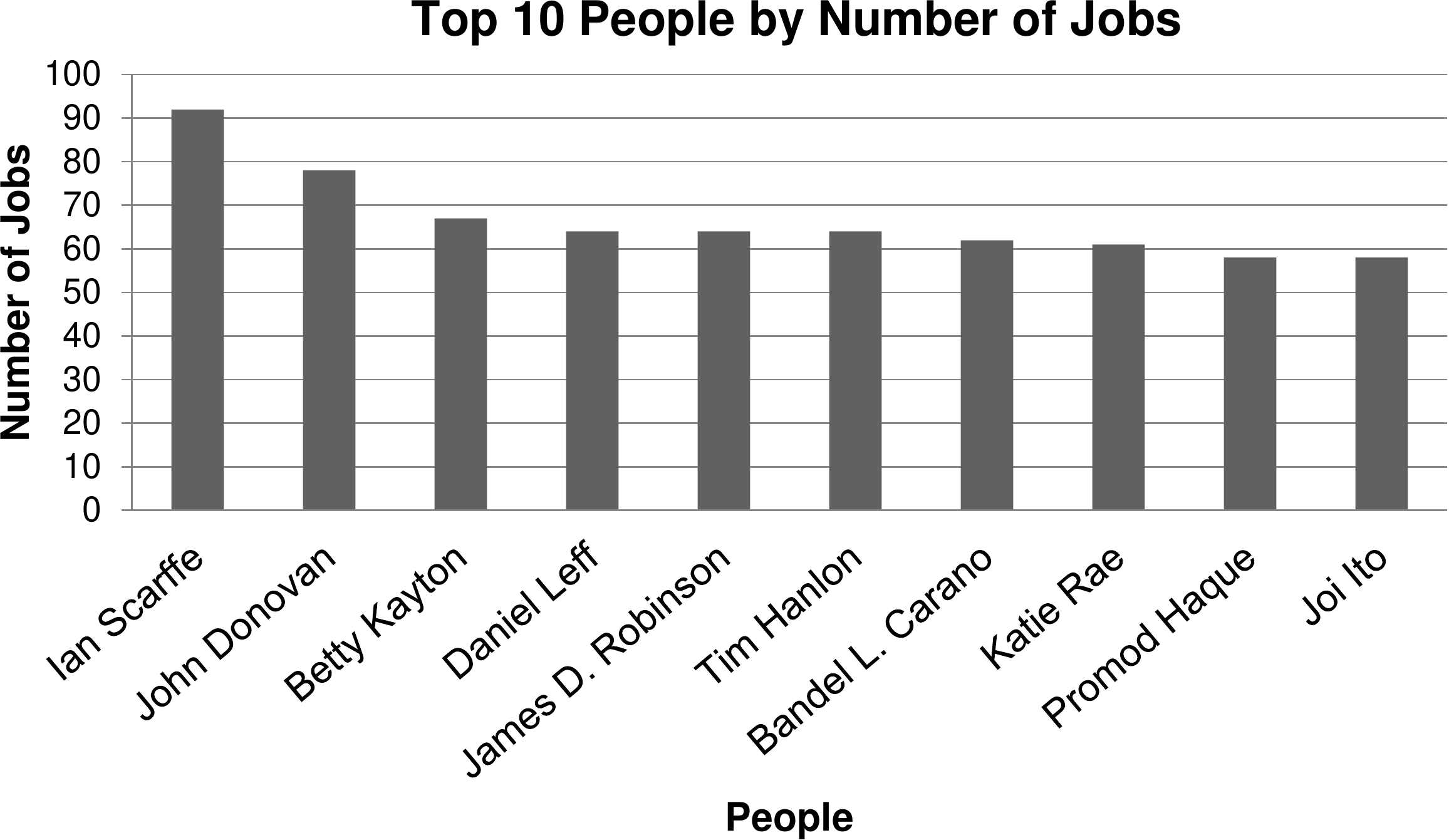}
 \end{subfigure}\hfil 
\begin{subfigure}{0.49\textwidth}
\raisebox{0.05cm}{
    \includegraphics[width=\linewidth]{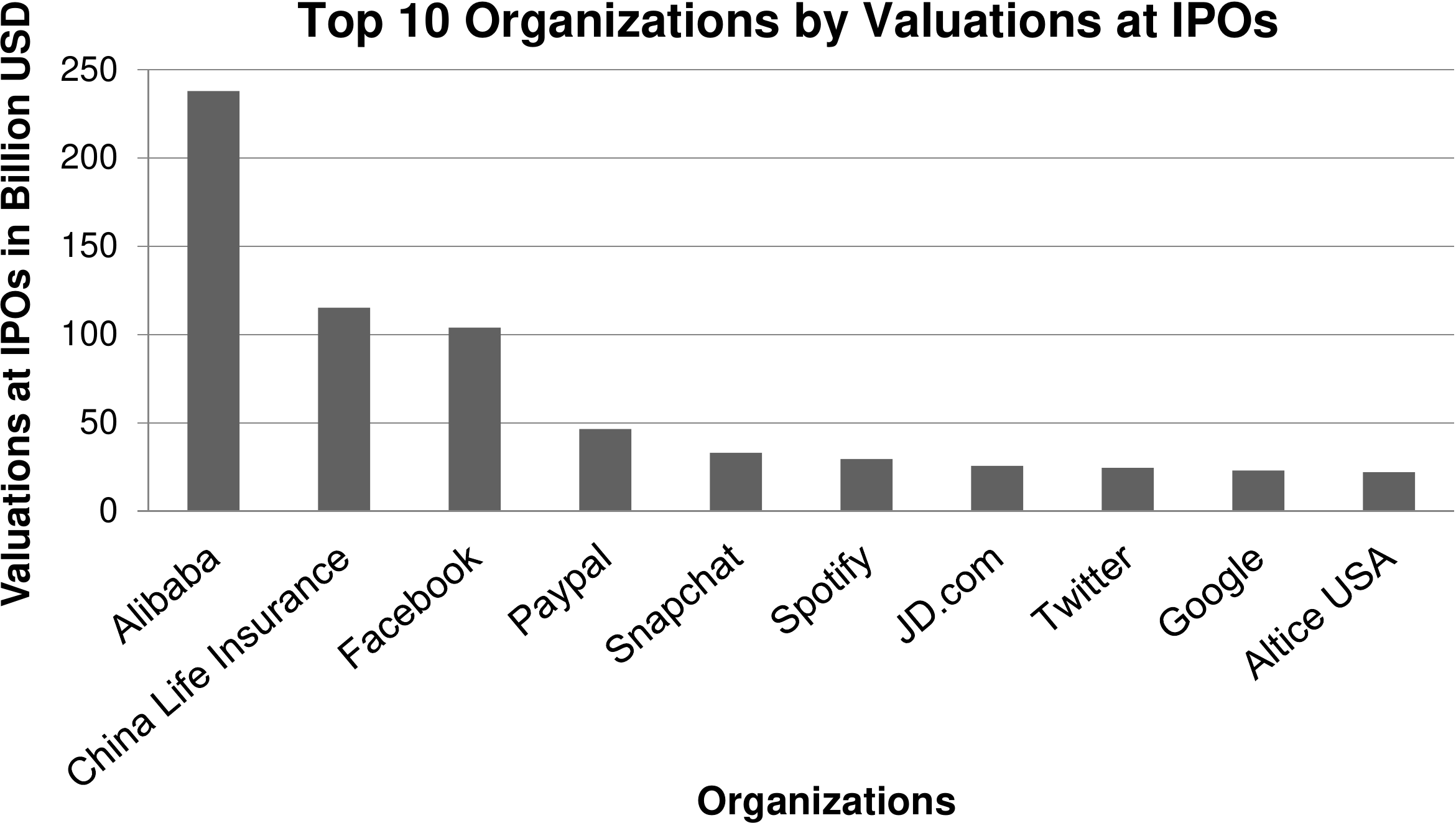}
    }
 \end{subfigure}\hfil 
\end{figure}

A Crunchbase RDF data set can be used for satisfying complex information needs by executing single SPARQL queries. Other data models and data processing methods, in contrast, may require high workload overhead and skills to transform the data into an appropriate format and query it efficiently. This can be exemplified by the figures in Fig.~\ref{fig:example-visualization}, which are generated based on single SPARQL queries. 
Lst.~\ref{lst:query1} and Lst.~\ref{lst:query2} show example SPARQL queries for retrieving the birth dates of investors stored in Crunchbase and for retrieving the continents ranked by the number of female CEOs, respectively.

\begin{lstlisting}[captionpos=b,label={lst:query2},caption={Querying the continents ranked by the number of female CEOs in organizations.},basicstyle=\scriptsize\ttfamily,frame=single,breaklines=true,float=t]
PREFIX rdf: <http://www.w3.org/1999/02/22-rdf-syntax-ns#>
PREFIX cb: <http://ontologycentral.com/2010/05/cb/vocab#>
PREFIX xsd: <http://www.w3.org/2001/XMLSchema#>

SELECT ?country COUNT(DISTINCT ?person) AS ?freq
FROM <http://linked-crunchbase.org>
WHERE {
 ?person rdf:type cb:Person .
 ?person cb:gender "Female" .
 ?person cb:jobs ?job .
 ?job cb:title "CEO" . 
 ?person cb:primary_location ?location .
 ?location cb:continent ?country .
}
GROUP BY ?country
ORDER BY DESC(?freq)
\end{lstlisting}

\subsection{Linked Data Set Descriptions and Ratings}

To follow the Linked Data best practices, we described our ontology by an OWL file, including Vocabulary-as-a-Friend (VOAF) descriptors. We also created a VoID file\footnote{See \url{http://www.w3.org/TR/void/} (Accessed: 18 July 2019).} for representing metadata about the generated RDF data set.\footnote{See \url{http://linked-crunchbase.org/void.ttl} (Accessed: 18 July 2019).}

We can classify our data set according to two 5-star rating schemes in the Linked Data context: 
\begin{enumerate}
   \item \textit{The 5-star deployment scheme for Open Data developed by Tim Berners-Lee:}\footnote{See \url{http://5stardata.info/} (Accessed: 18 July 2019).} Our Crunchbase RDF data set is a 5-star data set according to this scheme, because we provide our data set in RDF (leading to 4 stars) and link entity URIs 
   to DBpedia and our vocabulary URIs to other vocabularies (leading to 5 stars).
  \item \textit{Linked Data vocabulary star rating} \cite{Janowicz2014}: 
    This rating is intended to rate the use of vocabulary within Linked (Open) Data. 
    By providing an OWL-file, linking our vocabulary to \texttt{schema.org}, and creating a Vocabulary-of-a-Friend (VOAF) file,
we award the Crunchbase vocabulary 4 stars. 
\end{enumerate}
RDF data sets which are created in the future according to the procedure outlined in Sec.~\ref{sec:creating-data-set} can be attributed the same ratings.

\section{Usage}
\label{sec:usage}

The presented Crunchbase Linked Data API and Crunchbase RDF data are useful in a variety of scenarios as Crunchbase provides data which -- in large parts -- is not covered by other Linked Open Data (LOD) data sets. 
Furthermore, it becomes apparent that the Crunchbase RDF data set is useful not only to business people, but also to researchers of various fields, such as social studies and economics.


\subsection{Usage of the Crunchbase Linked Data API}

The following RDF wrappers for Crunchbase have been developed and used so far:
\begin{itemize}
 \item \textit{Semantic Crunchbase}\footnote{See \url{http://bnode.org/blog/2008/07/29/semantic-web-by-example-semantic-crunchbase} and the dedicated host \url{http://cb.semso.org/}  (Accessed: 18 July 2019), which is not available any more.} 
is an RDF wrapper for Crunchbase, released by Nowack shortly after the release of the official Crunchbase JSON API in July 2008.\footnote{See \url{http://techcrunch.com/2008/07/15/crunchbase-now-has-an-api-so-grab-our-data/} (Accessed: 18 July 2019).}
This initial Crunchbase wrapper transformed JSON provided by the Crunchbase API into RDF. However, no other data (such as \texttt{owl:sameAs} links) had been included and no external vocabulary (such as RDF, RDFS, or FOAF) had been used. The API is no longer available, but it shows that early efforts had been made to provide Crunchbase data in RDF.

 \item Harth et al.~\cite{Harth2013}
demonstrated in 2013 an on-the-fly integration of static and dynamic linked data sources. 
Among the data sources, an RDF version of Crunchbase was integrated to include office locations of technology companies in the overall system.
Harth et al. used an initial version of the Crunchbase Linked Data API presented in this paper. 

 \item In 2015, we built a first version of a Crunchbase RDF API and a Crunchbase RDF data set \cite{Faerber2018Crunchbase}. However, this RDF API is now outdated after updates on the official Crunchbase API and is no longer available. 
\end{itemize}

\subsection{Usage of the Crunchbase RDF Data Sets}

RDF data sets generated from the Crunchbase database have been used in the following ways so far:

\begin{itemize}
 \item Lee et al.~\cite{Lee2014}
presented an initiative of using linked data for financial data integration. 
The authors 
showed that the integrated RDF data allows a better comparison of financial reports, that it supports new KPI definitions, and that it allows timely access to external data.
Regarding Crunchbase, RDF data about funding, competitors, company acquisitions, main people in charge, and products were integrated into the framework. The authors of~\cite{Lee2014} used a first version of our Crunchbase RDF data set. In the current article, we present an updated version of the Crunchbase RDF data set, 
which contains 
considerably more entities and  more diverse entity types.

 \item In~\cite{Faerber2016},  
the Crunchbase RDF dump as presented in \cite{Faerber2018Crunchbase} was used for the purpose of monitoring news in order to find statements which are not in the Crunchbase knowledge graph yet. 
The \texttt{owl:sameAs}-relations between Crunchbase entities and DBpedia entities are used in order to apply an existing entity linking method for linking mentions in text to Crunchbase entities. 
Note that this RDF data set is published under CC-BY license, but is from 2015 and is thus outdated to a considerable degree. Furthermore, the data set only covers data about organizations, people, acquisitions, and directly linked entities. In the data set presented in this paper, in contrast, we cover for the first time all Crunchbase entities in RDF (347 million triples and 16 entity types) and also provide statistical key values of this data set.

\end{itemize}

\subsection{Further Potential Usage}

We can think of various use cases in which our Crunchbase API and RDF data set can be used.
First of all, as a very large data set in the domain-specific setting of tech companies and startups, covering information about investments, investors, and acquisitions, our Crunchbase data set can be used as a profound database for knowledge discovery and data mining methods. For instance, business analysts may receive answers to queries relating to investors' or investments' performance.\footnote{See ``Which companies in the category ``Semantic Web'' have got funded since 2000?'' in the Crunchbase mailing list post available at  \url{https://groups.google.com/d/msg/crunchbase-api/xiAQdg5CAo4/GN51XIlptWMJ} (Accessed: 18 July 2019).}
But, social analysis studies are also possible \cite{Xiang2012,Liang2016}. For instance, Liang and Yuan~\cite{Liang2016} use 12,000 companies and 12,000 people as data base for exploring how the similarity between investors and companies affects investing behavior. Our proposed data set contains about 780,000 people, 659,000 organizations, and various other information (e.g., news, jobs, websites, addresses, and investments) and as such can be used for experiments which either prove or refute the findings of such studies, or which allow completely new research questions to be answered. 

The full potential of our data set is likely to be unleashed when it is combined with other data, particularly with other RDF data sources. For instance, Crunchbase' information about the location and the technology sector of companies can be combined with information about job offers from an online job seeker platform \cite{Mochol:2007}. The data integration allows users to pose queries
 such as: ``Find all companies within the area of X which offer jobs in the field of Y.''
 
For dynamic monitoring of news and for further market monitoring purposes, Semantic Web methods such as text annotation (i.e., linking mentions in a text to their corresponding knowledge graph 
 entries) and relation extraction (i.e., extracting triples from text) are available. 
 However, these methods often only work well for specific, non-domain-specific underlying knowledge graphs, such as DBpedia. 
Since we provide also \texttt{owl:sameAs} links to DBpedia, we can still use the text annotation methods on the one hand and the Crunchbase data with its rich knowledge about innovative companies on the other.

\section{Conclusions}
\label{sec:conclusions}

In this paper, we proposed a Crunchbase Linked Data API based on the Crunchbase JSON API. This API is available online at \url{http://linked-crunchbase.org}. 
Furthermore, we described our method to crawl data with this Linked Data API and created a custom Crunchbase RDF knowledge graph. 
To ensure the best possible usage and impact of the Linked Data API and of RDF data sets obtained by using the API, we proceeded along the Linked Data best practices. This included (1)~a description of the API, the RDF dump, and the schema via published OWL and VoID files, (2)~a mapping of Crunchbase relations and classes to relations and classes from other vocabularies, and (3)~the integration of \texttt{owl:sameAs} links to entities in DBpedia.

In the future, 
we intend to improve the linkage to DBpedia and to create links to other linked data sources. 
Furthermore, we plan to investigate 
how often and in which way companies, key people, and investors are mentioned over time in news articles using Crunchbase as an underlying knowledge graph.

\bibliographystyle{splncs}
\bibliography{paper}

\end{document}